\documentclass[aps,pre,twocolumn,showpacs]{revtex4}

\usepackage{natbib}
\usepackage{graphicx}
\usepackage{amsmath}
\usepackage{color}
\usepackage{url}
\usepackage{subfigure}
\usepackage{dcolumn}
\usepackage{bm}

\DeclareMathOperator{\Deg}{Deg}

\DeclareMathOperator*{\argmax}{argmax}
\DeclareMathOperator*{\origin}{\mathcal{O}}

\begin{document}


\title{Towards real-time community detection in large networks}

\author{Ian X.Y. Leung}
\email{[firstname.lastname]@cl.cam.ac.uk}
\author{Pan Hui}
\email{[firstname.lastname]@cl.cam.ac.uk}
\author{Pietro Li\`{o}}
\email{[firstname.lastname]@cl.cam.ac.uk}
\author{Jon Crowcroft}
\email{[firstname.lastname]@cl.cam.ac.uk}
\affiliation{Computer Laboratory, University of Cambridge, Cambridge CB3 0FD, U.K.}


\begin{abstract}
The recent boom of large-scale Online Social Networks (OSNs) both enables and necessitates the use of parallelisable and scalable computational techniques for their analysis. We examine the problem of real-time community detection and a recently proposed linear time---$O(m)$ on a network with $m$ edges---label propagation or ``epidemic" community detection algorithm. We identify characteristics and drawbacks of the algorithm and extend it by incorporating different heuristics to facilitate reliable and multifunctional real-time community detection. With limited computational resources, we employ the algorithm on OSN data with 1 million nodes and about 58 million directed edges. Experiments and benchmarks reveal that the extended algorithm is not only faster but its community detection accuracy compares favourably over popular modularity-gain optimization algorithms known to suffer from their resolution limits. \end{abstract}

\maketitle

\section{Introduction}
Recent years have seen the flourishing of numerous Online Social Networks (OSNs). Cyber communities such as Facebook, MySpace and Orkut, where users can keep in touch with friends on the Internet, have all emerged as top 10 sites globally in terms of traffic. Tools and algorithms to understand the network structures have consequently emerged as popular research topics. By their nature, OSNs contain an immense number of person nodes which are sparsely connected. Edges are often bidirectional since a mutual agreement is required before such friendship links are established. One of the most notable phenomenon in such networks is the resemblance of the so-called 6-degree of separation \cite{xxx} where on average every person is related to another random person via 5 other people in the real world. This has indeed been shown in real life communities and, much more conveniently, on online communities \footnote{See the Facebook.com Six Degrees Project.}. Networks which exhibit such small degrees of separation while being sparsely connected are famously known as Small-World Networks \cite{swn}.

\begin{figure}
\begin{center}
  \includegraphics[trim = 160 40 150 10, clip, width=\linewidth]{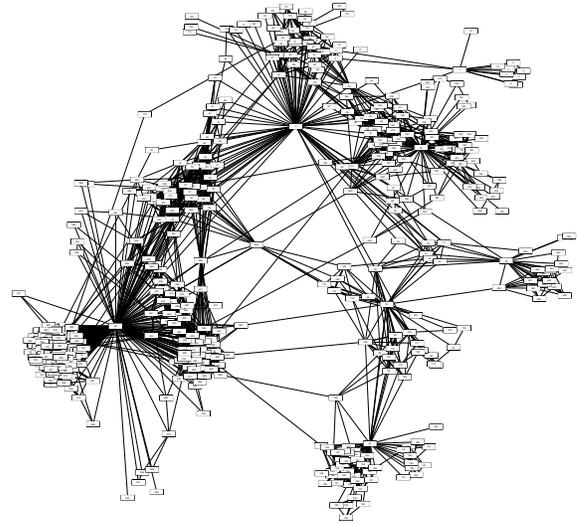}
\caption{\label{osn} Snapshot of a subgraph of an OSN (500 nodes).}
\end{center}
\end{figure}

Well established online communities often contain tens of millions of users connected by some billions of edges which enable---and necessitate---the use of parallelisable and scalable computational techniques for their analysis. In this literature, we examine the problem of network community detection. Graphically, such communities are characterized by a group of nodes which are densely connected by internal edges but less so towards the outside of the communities, as depicted by the densely connected subgraphs in Fig. \ref{osn}. Understanding the community structure and dynamics of networks is vital for the design of related applications, devising business strategies and may even have direct implications on the design of the networks themselves \cite{mislove}.

We empirically analyse a recently proposed community detection technique by label propagation discussed in \cite{linear}, which is summarised as follows. Each node in a network is first given a unique label. Every iteration, each node is updated by choosing the label which most of its neighbours have (the maximal label). If there happens to be multiple maximal labels (which is typical in the beginning), one label is picked randomly. Previous results have shown that this algorithm is extremely efficient in uncovering accurate community structure. As an example, we apply the algorithm on a set of OSN connection data crawled by Mislove et al. \cite{mislove} of 3 million nodes connected by roughly 0.2 billion directed links.

We give a survey of related work in the next Section and look further into the characteristics of the algorithm in Section \ref{diss}. We discuss the potential implementations, improvements and applications of the algorithm on different types of networks (Section \ref{imp}). Section \ref{comp} gives detailed comparisons between the label propagation algorithm (LPA) and fast modularity-optimization algorithms. We conclude the paper with future directions of research in Section \ref{con}.

\section{Related Work}\label{related}
Community detection in complex networks has attracted ample attention in recent years. Apart from OSNs, researchers have engaged in community analysis in various types of networks. In the case of the Internet, examples of communities are found in autonomous systems \cite{lusseau-2004-271} and indeed web pages of similar topic \cite{flake02self-organization}. In biological networks, it is widely believed that modular structure plays a crucial role in biological functions \cite{citeulike:302050}. Related literatures such as \cite{newman04detectCommunity,danon-2005,latora} may serve as introductory reading, which also include methodological overviews and comparative studies of different algorithms.

The detection of community structure in a network is generally intended as a procedure for mapping the network into a tree \cite{radicchi-2004-101}, known as dendrogram. In this tree, the leaves are the nodes and the branches join them or (at a higher level) groups of them, thus identifying a hierarchy of communities. Nodes can either be agglomerated successively starting from single nodes (agglomerative), or the whole network can be recursively partitioned (divisive). Newman and Girvan introduced a seminal divisive algorithm in which the selection of the edge to be cut is based on the value of its edge betweenness \cite{newman04findCommunity}, the number of shortest paths between all node pairs running through it. It is clear that when a graph is made of tightly bound clusters, each loosely interconnected, all shortest paths between nodes in different clusters have to go through the few inter-cluster connections, which therefore have a large betweenness value. Recursively removing these large betweenness edges would partition the network into communities of different sizes.

Quantitatively, however, we need a metric to measure how well the community detection is progressing, otherwise most algorithms would either continue until every node is split into a single community or all join together into one. Newman and Girvan proposed in \cite{newman04findCommunity} a measure of the goodness of communities called \textit{modularity}, for the set of uncovered communities $\mathcal{C}$, the modularity is defined to be :
\begin{equation}
Q = \sum_{c \in \mathcal{C}} \left(\frac{I_c}{E} - \left(\frac{2I_c + O_c}{2E}\right)^{2}\right),
\end{equation}
where $I_c$ indicates the total number of internal edges that have both ends in $c$, $O_c$ is the number of outgoing edges that have only one end in $c$ and $E$ is the total number of edges. This measure essentially compares the number of links inside a given module with the expected value for a randomized graph of the same size and same degree sequence.

The concept of modularity has gained such popularity that it has not only been used as a measure of the community partitioning of a network but also as a key fitness indicator in various community detection algorithms. The algorithm proposed by Clauset, Newman and Moore (CNM) \cite{clauset-2004-70}, which greedily combines nodes/communities to optimize modularity gain, is perhaps to date one the most popular algorithms in detecting communities in relatively large scale networks. In the time when CNM was proposed, it was then the only algorithm capable of community detection on networks of size 500,000 in a matter of hours. Throughout the years, several variations of the CNM have been proposed \cite{dda, wak07, lam08}. Most of them concentrate on more efficient data structures as well as modularity gain heuristics to improve the overall performance. A latest adaptation \cite{lam08} that treats newly combined communities as a single node after each iteration is able to identify community structure on a network containing 1 billion edges in a matter of hours.

It is vital, however, to understand that modularity is not a scale-invariant measure and hence, by blindly relying on its maximization, detection of communities smaller than a certain size is impossible. This is famously known as the resolution limit \cite{reslimit} of modularity based algorithms. Since LPA does not involve modularity optimization, its community detection capability is scale-independent and therefore not affected by the resolution limit as will be shown in Section \ref{comp}.

\section{Discussion}\label{diss}
Here, we give a brief discussion on the characteristics of the algorithm as well as some preliminary results applying the algorithm on the OSN described above.

\subsection{A ``near linear time" algorithm}
One can consider the label spreading as a simplified but specific case of epidemic spreading where all individuals are considered infectious with their own unique disease. Each person is infected by a disease that is prevalent in his or her neighbourhood. Fig. \ref{label} depicts the labelling convergence seen in a $4$-clique. The number of clusters monotonically decreases each iteration as certain labels become extinct due to domination by other labels. With certain rare and exceptional cases, the labelling self-organises to an unsupervised equilibrium efficiently.

\begin{figure}[htb]
\begin{center}
  \includegraphics[width=\linewidth]{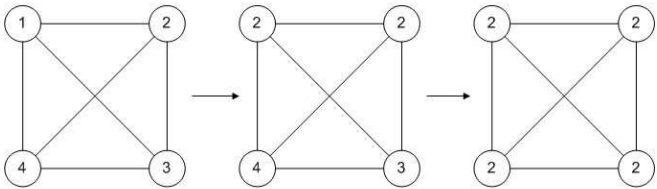}

  \caption{\label{label} Each node is looked at in a certain order and a new label is selected. The above shows how nodes in a $4$-clique self-organise into one single community in one iteration.}
   \end{center}
 \end{figure}

As suggested in \cite{linear}, certain properties may prevent the equilibrium from occurring. For instance, a network with a bipartite structure might render the system to oscillate if the algorithm is run \emph{synchronously}, i.e., all nodes are updated together only after they have selected their maximal labels. Running the algorithm \emph{asynchronously} in a randomized order every iteration, as suggested in the paper, may result in less definitive results but solves the problem. It was also suggested that a node that has two equally maximal labels to choose from may fail to converge and an extra stopping criterion to prevent the switching of label would have to be in place. It is, however, noted in our implementation that including the concerned label itself into the maximal label consideration effectively avoids all the above non-convergent behaviours and the requirement for an extra stopping criterion.

In one iteration, each node's neighbours are examined and the maximal label is chosen. The running time of this algorithm is therefore $O(knd)$, where $k$ is the number of iterations, $n$ the number of nodes and $d$ the average degree of nodes. Note that $nd$ can also be described by $m$, the number of edges. The number of iterations required, $k$, is dependent on the stopping criterion but is not very well understood. \cite{linear} suggested that the number of iterations required is independent to the number of nodes and that after 5 iterations, 95\% of their nodes are already accurately clustered.

\begin{figure}[htb]
\begin{center}
  \includegraphics[trim = 0 30 20 30, clip, width=\linewidth]{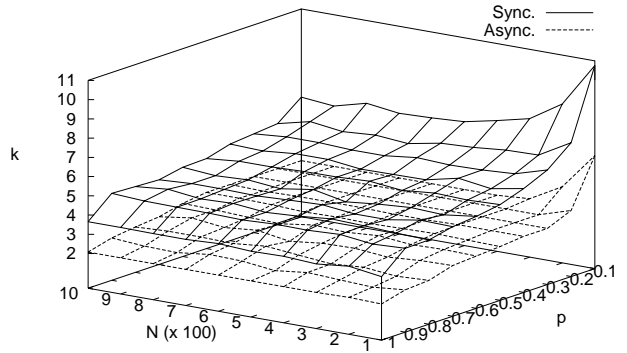}
  \caption{\label{clique}The above plots show the number of iterations required before convergence for both the synchronous and asynchronous implementations on a random graph of size $N$ with probability of pairwise connection $p$. All values here are averaged over 100 realisations.}
\end{center} \end{figure}

Since labels can hardly affect nodes outside their local densely connected substructures, the convergent behaviour should be dependent on these substructures rather than the whole network. This is confirmed by preliminary testing and directs us to look at substructures which can ultimately become the community. Experiments show that the average number of iterations required for the labelling to converge (no change in labels) in an $N$-clique for the asynchronous and synchronous implementations are 2.1 and 3.6 respectively, highly independent of $N$. To further investigate the average convergent behaviour on a substructure, we look at Fig. \ref{clique} which summarises the relationship between number of iterations required before convergence, $k$, to the pairwise connectivity, $p$, that controls the edge density in a random graph of size $N$ (where $p = 1$ corresponds to the $N$-clique).

In both implementations, we see that $k$ remains fairly constant over both $N$ and $p$ until $p$ reaches a certain threshold, which when reached we begin to see an inverse dependence between $N$ and $k$. The overall averages of asynchronous and synchronous implementations in this case are 2.8 and 5.2.

\begin{figure}[htb]
\begin{center}
  \includegraphics[trim = 130 10 180 10, clip, width=0.8\linewidth]{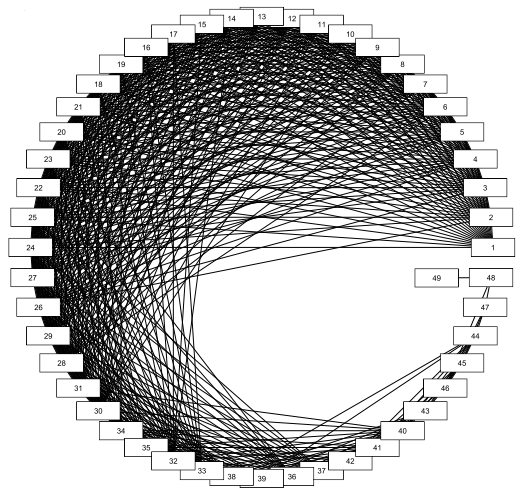}
  \caption{This substructure is constructed on an $N$-clique, $N = 25$, by attaching each new node, labelled $l$ ,$N<l<2N$, to existing nodes $l-1\ldots2(l-N)$, thus contains 49 ($2N-1$) nodes and 600 ($N(N-1)$) edges.}\label{25clique}
  \end{center}
\end{figure}

Let us, however, consider another simple but non-random topology. Suppose we start off with an $N$-Clique, at each $j^{th}$ construction, the graph is grown by connecting the $N-j$ most recently joined nodes to the new node (c.f. Fig. \ref{25clique}).

\begin{figure}[htb]
\begin{center}
  \includegraphics[width=\linewidth]{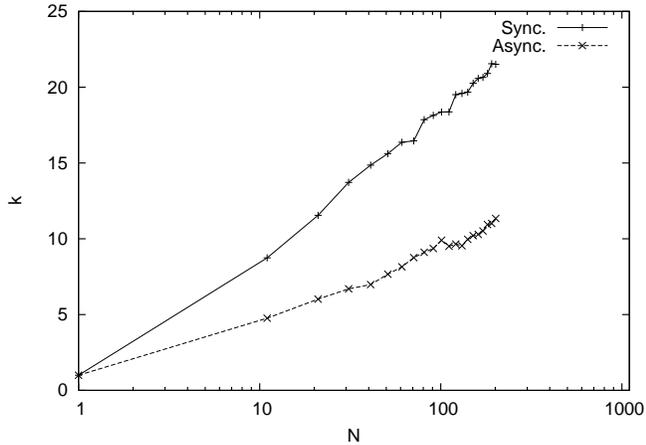}
  \caption{The relationships between the number of iterations required before convergence, $k$, of both implementations to the size, $N$, of the aforementioned structure. All values here are averaged over 100 realisations.}\label{cl}
\end{center} \end{figure}

These structures by construction will converge into a single community by LPA. Without worrying about how abundant such patterns are in real world communities, we look at the convergent behaviour shown in Fig. \ref{cl}.
The trend clearly reveals that $k$ grows logarithmically with respect to $N$. We therefore suggest the possible worst case of $k$ of the order of $O(\log{N})$, where $N$ is the size of the largest substructure with a topology similar to the above. Indeed, we anticipate real world social networks to contain highly heterogeneous substructures which may be intricately connected to affect each other's convergence. We thus consider the understanding of the convergent behaviour in large complex networks such as OSNs as a direction for further investigation.

\subsection{Community Detection in OSN}
\begin{figure}[htb]
\begin{center}
  \includegraphics[width=\linewidth]{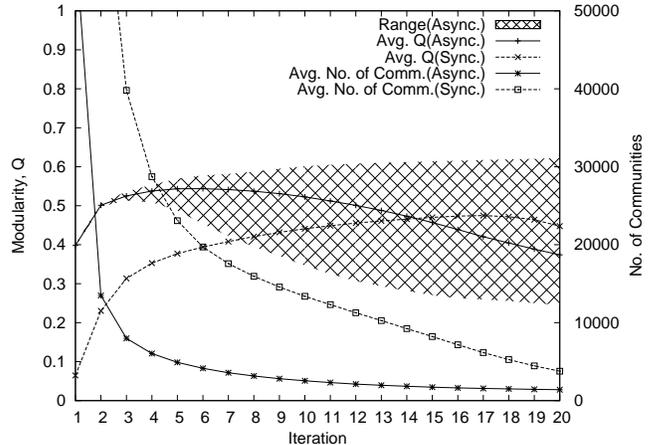}
  \caption{\label{f6}Average performances of asynchronous and synchronous LPA. Values are averaged over 5 Runs. Shaded area denotes the range of the performances of asynchronous implementation.}
\end{center} \end{figure}

We carry out community detection on the aforementioned OSN using a desktop PC with 4GB ram and a 2.4 GHz quad-core processor running 32-bit Java VM 1.6. Due to limited memory, we restrict the number of nodes to the first million. Since the order of nodes in the original data corresponds to that of a breath-first web crawling, this way of ``cutting off" the data is equivalent to extracting a snowball sample. As discussed in \cite{mislove}, snowball methods are known to over-sample high-degree nodes, under-sample low-degree ones and overestimate the average node degree. This is seen by the higher average degree of the subgraph, 250, compared to 106 of the original graph. Nonetheless, since the purpose of this literature is to evaluate the algorithm on large-scale networks, the sampled network satisfies our requirements. The sampled subgraph contains 1,000,000 nodes and 58,793,458 directed links. Convergent behaviours of the two different implementations are shown in Fig. \ref{f6}.

A crucial point is that in a complex network as large as this, the so called ``convergence" does not necessarily yield an optimal result in terms of modularity. For example, we see the asynchronous implementation merely took on average 5 iterations to achieve a maximum modularity but has highly volatile results in different runs as depicted by the shaded area in the figure. On the other hand, the synchronous implementation achieved maximum modularity much slower than the asynchronous version but its performance on average is much more stable (its performance range is thus omitted). The performances of these two different implementations are equally important to be understood and utilised. Further discussions on the implications of these implementations and their utilizations are given in Section \ref{imp}.

Each single-threaded iteration finishes in a matter of tens of second and thus, depending on the stopping criterion, it can take as little as 8 to 10 minutes up to peak performance. Extrapolating the time required with respect to the number of edges, the algorithm without any optimization should be able to detect communities on a graph with 1 billion edges in less than 180 minutes, in a magnitude similar to that in \cite{lam08}.

\begin{figure}[htb]
\begin{center}
  \includegraphics[width=\linewidth]{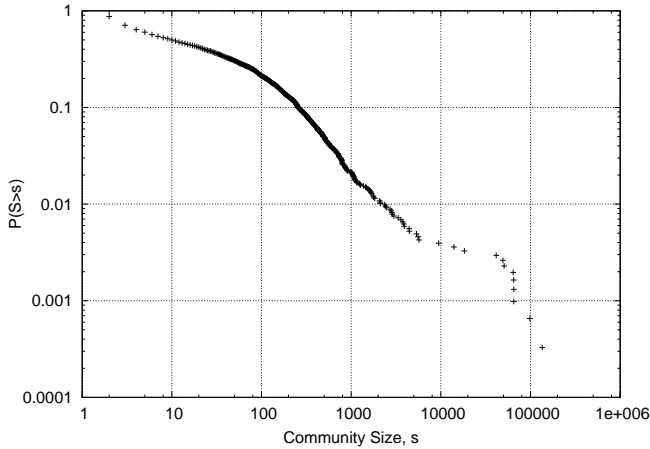}
  \caption{The community-size distribution of communities uncovered by the algorithm, which follows a 2-part power law.}\label{comm}
\end{center} \end{figure}

Fig. \ref{comm} shows the distribution of community/cluster size collected by a specific run of the asynchronous version of the algorithm when the modularity peaked at 0.638. The size distribution of communities within the OSN follows a 2-part power law distribution in the complementary CDF with an estimated coefficient of 1.1. The interested reader is referred to \cite{big, latora} for discussions on the characteristics of different networks.

\section{A more reliable and efficient algorithm}\label{imp}
In this section, we discuss potential modifications to the algorithm to increase its reliability, functionality and computational efficiency.

\subsection{Hop Attenuation \& Node Preference}\label{hop}
Due to the ``epidemic" nature of the algorithm, a major limitation of the algorithm is noted where certain ``label epidemic" manages to ``plague" a large amount of nodes. To be exact, in some runs a certain community of size over 500,000 (50\% of the number of nodes) is formed---as opposed to the thousand other counterparts averagely sized in a magnitude of 100s---greatly contributing to modularity drop after the peak. We conjecture that this is partially due to the asynchronous nature of the algorithm and the initial formation of communities, where certain communities do not form strong enough links to prevent a foreign ``epidemic" to sweep through. Further experiments confirm that the synchronous version of the algorithm slows down the formation of such ``monster" communities but do not prevent them.

We propose an extension to this algorithm by adding a score associated with the label which decreases as it traverses from its origin. A node is initially given a score of 1.0 for its label. After a node $i$ has collected from its neighbourhood, $\mathcal{N}_{i}$, all the respective labels and the scores, the calculation of the new maximal label, $\mathcal{L}'$, can be generalised by:

\begin{equation}\label{label}
    \mathcal{L}'_{i} = \argmax_{\mathcal{L}} \sum_{i' \in \mathcal{N}_{i}}{s_{i'}(\mathcal{L}_{i'}) \cdot {f(i')}^{m}} \cdot w_{i',i},
\end{equation}

where $\mathcal{L}_{i}$ is the label of node $i$, $s_{i}(\mathcal{L})$ is the hop score of label $\mathcal{L}$ in $i$, $w_{i',i}$ is the weight of the edge between $i'$ and $i$ (we sum the weights in both directions if the graph is directed) and $f(i)$ is any arbitrary comparable characteristic for any node $i$. For instance, if we define $f(i) = \Deg(i)$, when $m > 0$, more preference is given to node with more neighbours; $m < 0$, less. The final step is to assign a new attenuated score $s'$ to the new label $\mathcal{L}'$ of $i$ by subtracting hop attenuation $\delta$, $0 < \delta < 1$:

\begin{equation}\label{label2}
   s'_{i}(\mathcal{L}'_{i}) = \left(\max_{i' \in \mathcal{N}_{i}(\mathcal{L}'_{i})}{s_{i}(\mathcal{L}_{i'})}\right) - \delta,
\end{equation}

where $\mathcal{N}_{i}(\mathcal{L})$ is the set of neighbours of $i$ that has label $\mathcal{L}$. The value $\delta$ governs how far a particular label can spread as a function of the geodesic distance from its origin. This additional parameter adds in extra uncertainties to the algorithm but may encourage a stronger local community to form before a large cluster start to dominate. Ideally, the selection of $\delta$ can even be adaptive to current number of iteration, the neighbourhood of the node concerned and perhaps some \emph{a priori} network parameters. We investigate the use of varying $\delta$ in the next section and assume here a constant value for $\delta$. Note that this setting may induce a negative feedback loop, we therefore let $\delta = 0$ if the selected label is equal to the current label.

\begin{figure*}[htb]
\centering
Synchronous:\\
    \subfigure[]{\label{fm}\includegraphics[width=0.338\linewidth, height = 123pt]{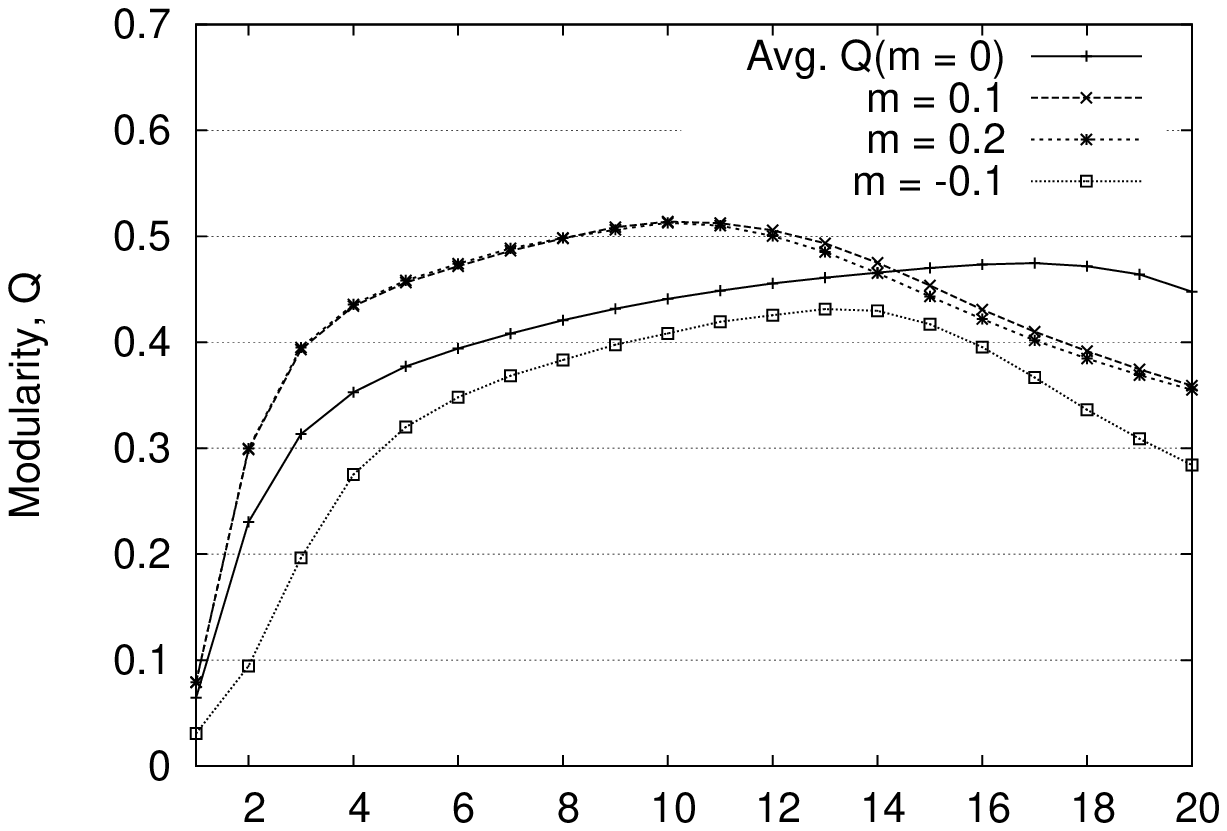}}
    \subfigure[]{\label{fd}\includegraphics[width=0.325\linewidth, height = 123pt]{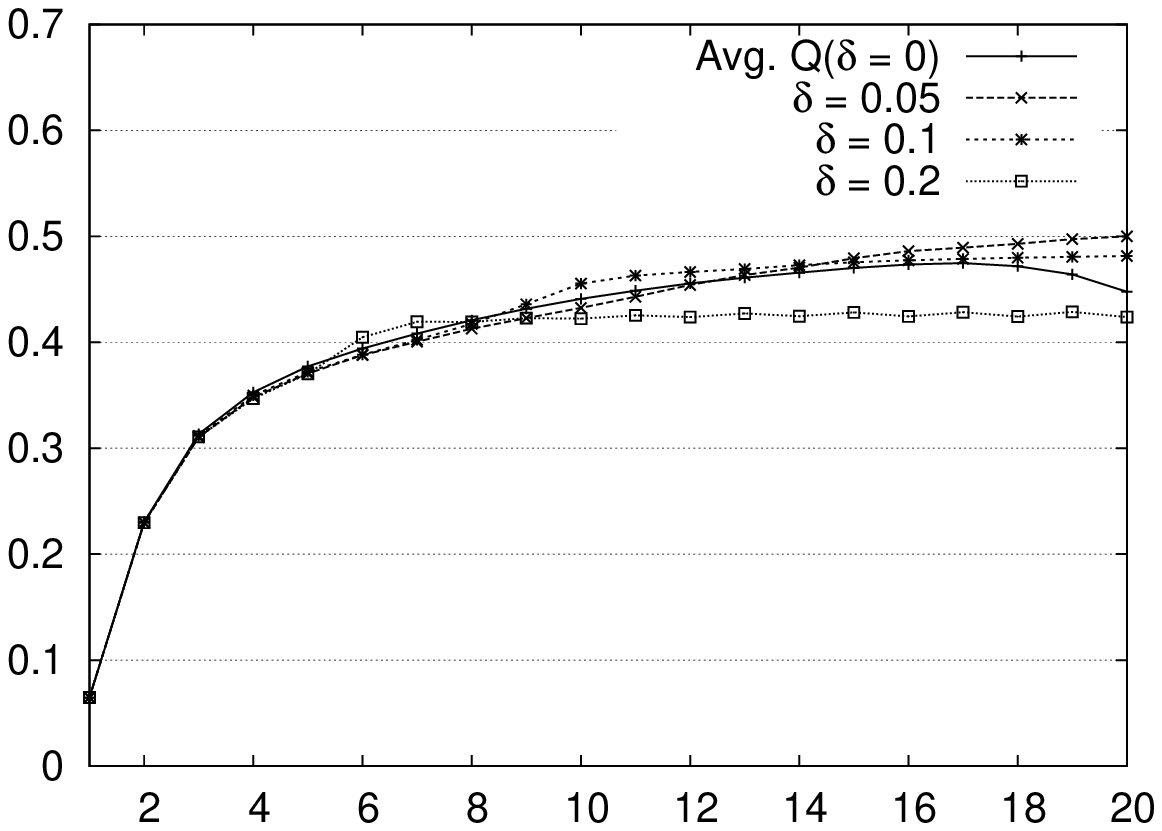}}
    \subfigure[]{\label{fx}\includegraphics[width=0.325\linewidth, height = 123pt]{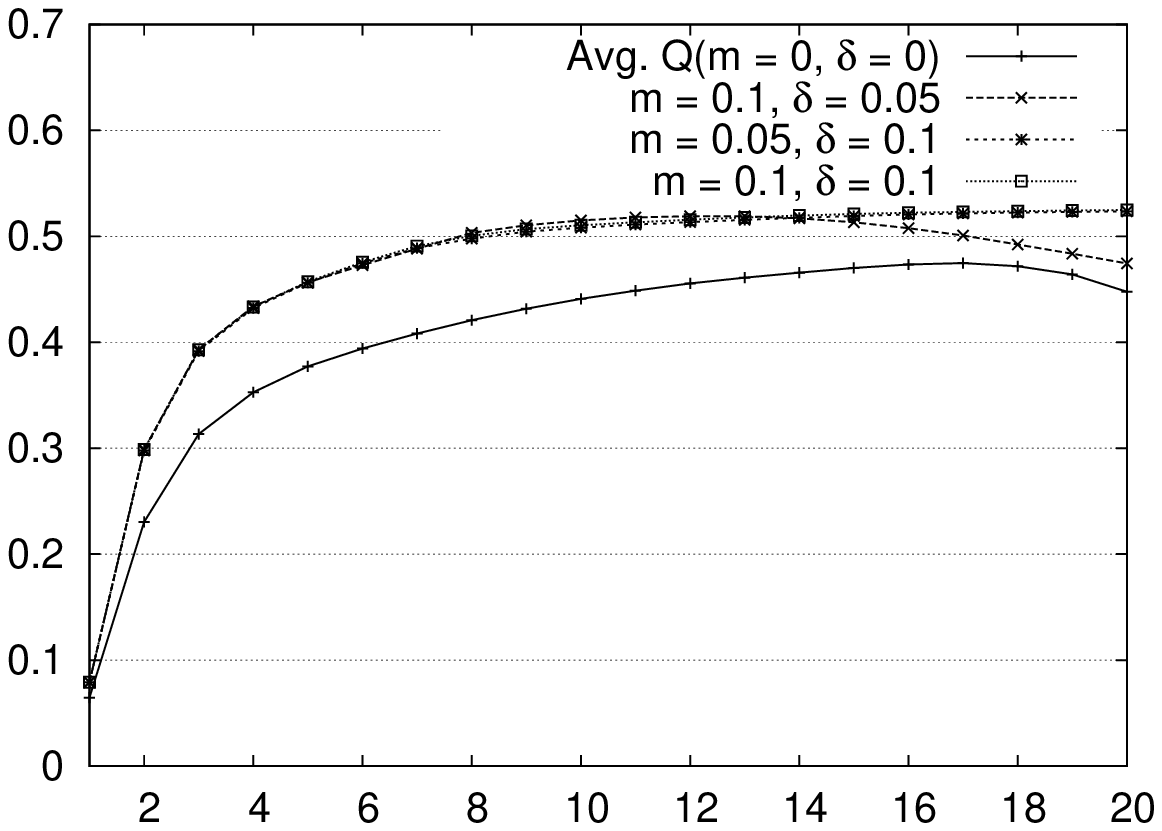}}\\Asynchronous:\\
    \subfigure[]{\label{fmf}\includegraphics[width=0.338\linewidth, height = 130pt]{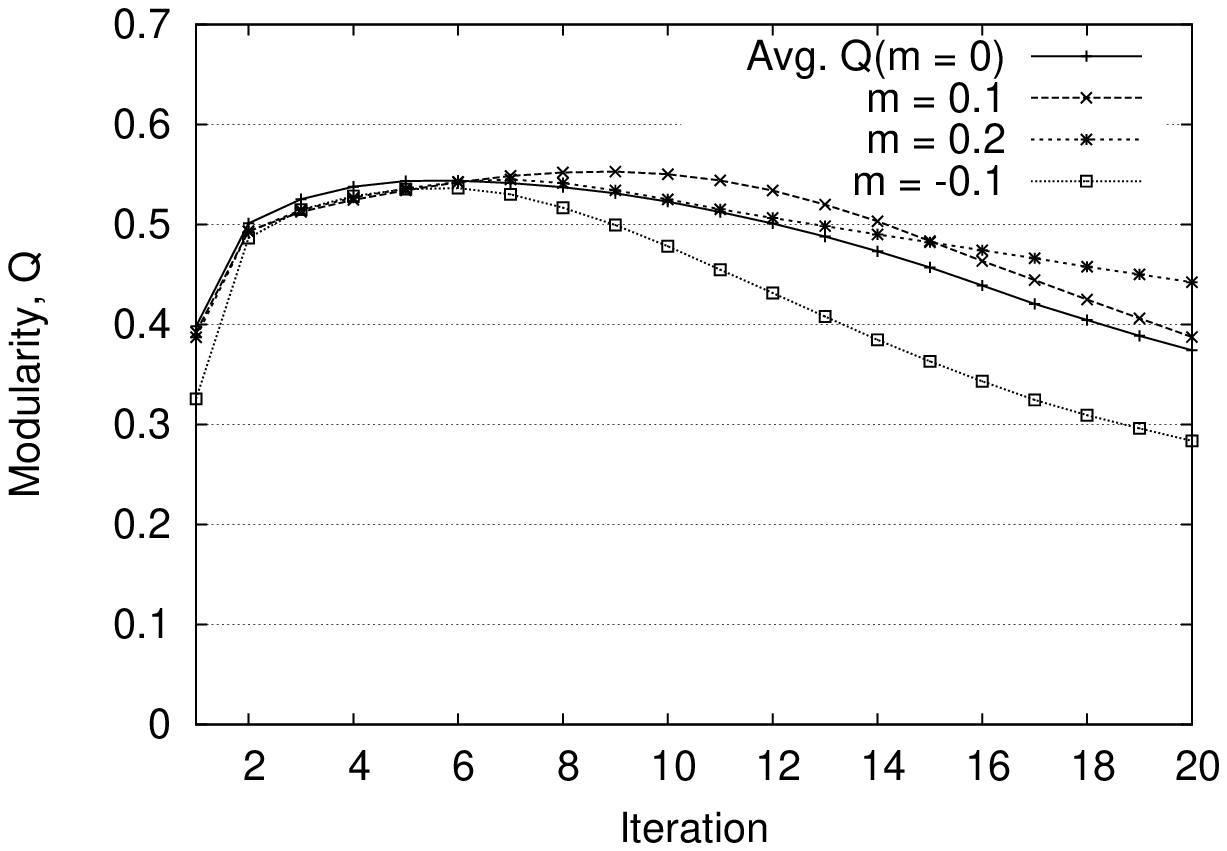}}
    \subfigure[]{\label{fdf}\includegraphics[width=0.325\linewidth, height = 130pt]{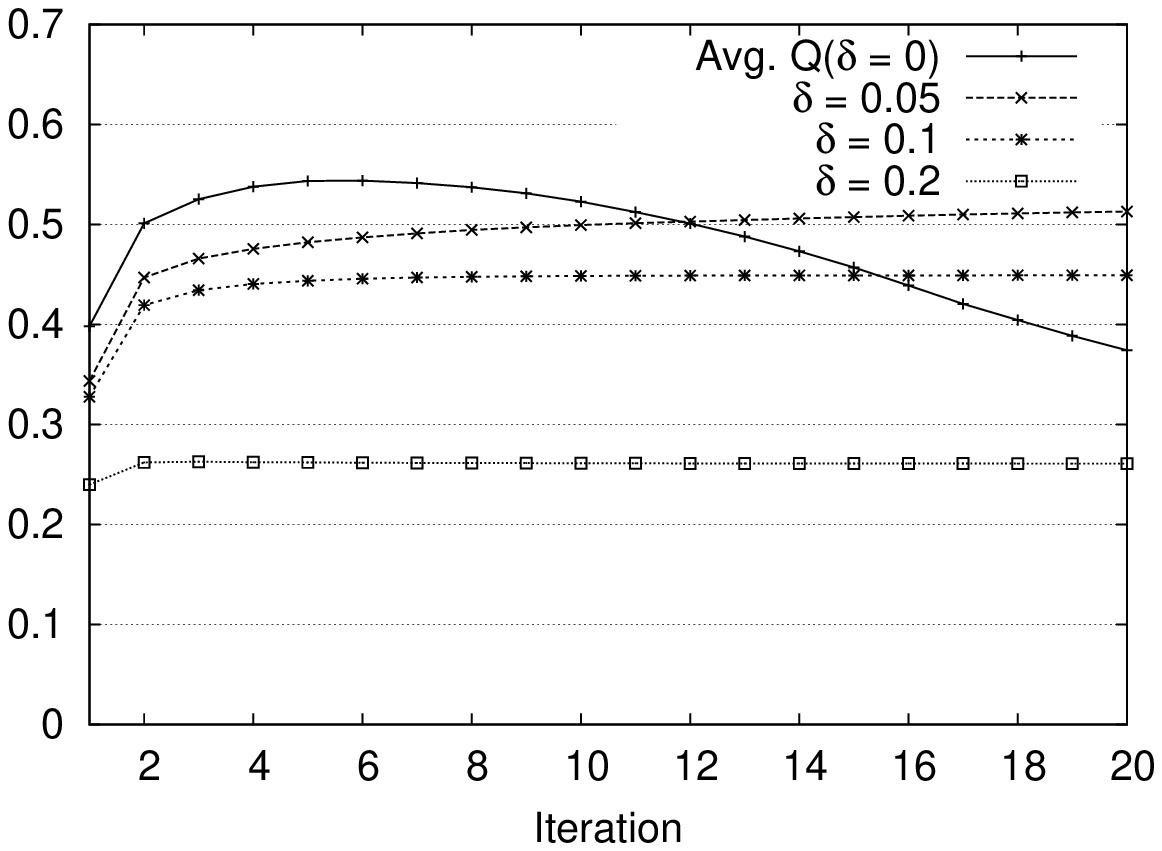}}
    \subfigure[]{\label{fxf}\includegraphics[width=0.325\linewidth, height = 130pt]{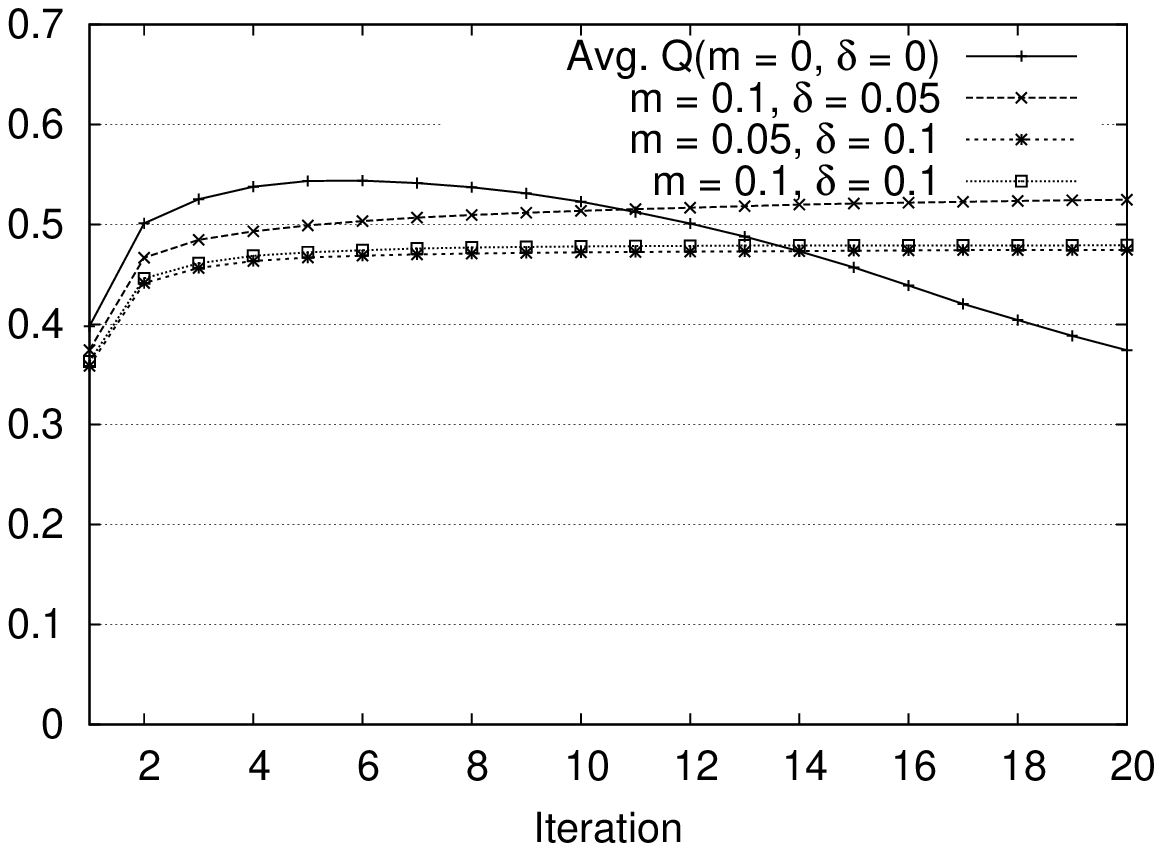}}
    \caption{Average performance comparisons of the synchronous and asynchronous implementations with varying $\delta$ and $m$ over 5 Runs.}
    \label{test4}
\end{figure*}

As discussed, modularity has been widely used in the literature as a metric to contrast the community detection capabilities on real world networks between different algorithms. Whilst high modularity indicates a significant modularised structure over a randomised graph of the network concerned, the correspondence between high modularity and accurately partitioned communities is not well understood due to the resolution limit of modularity.  Here we attempt to contrast the behaviours of the algorithms on the OSN based on modularity but shall not draw strong conclusions on the accuracies of the community detection due to the above reasons. In Section \ref{comp}, a novel benchmark proposed by Lancichinetti et. al. \cite{bm} capable of revealing resolution limit of modularity-based algorithms is used for further comparisons.

Fig. \ref{test4} depicts the average performance curves over 5 runs for both versions of the algorithm applying hop attenuation and preferential linkage. The results suggest that, on both implementations, a slight but not too high a preference on high-degree nodes ($m > 0$) can speed up the process for achieving peak modularity on the OSN network but also gives rise to a steeper drop as shown in Fig. \ref{fm}. We believe, however, different magnitudes of $m$ simply restrict the choice of nodes to different subsets, some of which may contribute to a ``global pandemic" and some may not. By simply using the degree of a node may not be a heuristic generic enough for different networks. Further study is required to understand, if at all possible, how to deduce a generic preference on neighbourhood labels every iteration without resorting to a global metric, which is costly. Nonetheless, we show that giving preference to certain nodes over others when deciding between labels to accept can be beneficial in terms of number of iterations to achieve maximum modularity.

Looking at hop attenuation, we find that the application of $\delta$ indeed deters the occurrence of the ``monster clusters" as expected and thereby preventing the modularity drop after certain iterations. But it was also obvious that high hop attenuation prevented the healthy growing of the communities and restricted the increase in modularity (c.f. Fig. \ref{fd},\ref{fdf}). Moreover, we conjecture that hop attenuation restrains the spread of the label from an arbitrary center and thereby the formation of circular clusters. This suppression in forming non-circular clusters may lead to the suboptimal performance in terms of modularity, as shown in the asynchronous case (Fig. \ref{fdf}). Finally, from Fig. \ref{fx} and \ref{fxf}, we see that combining both parameters, on average, benefits both versions of the algorithm in achieving a community partitioning of high modularity more efficiently and consistently.

\subsection{Hierarchical \& Overlapping Communities}\label{hie}
Communities in certain networks are known to be hierarchical. For instance, students in the same classes often form some strong local communities while these communities, say of the same school, in turn form a larger but relatively weaker community. As discussed in Section \ref{related}, most CNM-based algorithms are inherently hierarchical since communities are agglomerated by greedy local optimization of modularity gain.

We present two simple modifications to the original method to enable the detection of hierarchical communities. Firstly, let us consider the application of hop attenuation on label propagation. Suppose we impose a very high hop attenuation at the beginning, we expect communities of small diameter to form. If we then gradually relax the attenuation value, we should expect these small communities to merge into larger ones. In order to achieve this, we modify eq. (\ref{label2}) as follows:

\begin{equation}\label{label3}
   s'_{i}(\mathcal{L}_{i}) = 1 - \delta(d_G(\origin(\mathcal{L}_{i}),i)) , \\
\end{equation}
where
\begin{equation}\label{label4}
   d_G(\origin(\mathcal{L}_{i}), i) = 1 + \min_{i' \in \mathcal{N}_{i}( \mathcal{L}_{i})}{d_G(\origin(\mathcal{L}_{i}),i')} .
\end{equation}

Essentially, instead of receiving the current hop scores from the neighbourhood and carry out a subtraction, the score is now determined by the actual geodesic distance ($d_G$) from the label $\mathcal{L}$'s origin, denoted by $\origin(\mathcal{L})$ and the function $\delta$. This gives greater flexibility of $\delta$ in terms of geodesic distances and can facilitate iteration-dependent hop attenuation as required here with slight extra computation cost.

Our second proposal is inspired from \cite{lam08}, where we can similarly treat newly combined communities as a single node, and use the number of inter-community edges as the weight of edges between these ``fresh condensed" nodes. Instead of doing this every iteration, we can apply certain amount of hop attenuation or hard limit in terms of the diameter of the community and do this after an equilibrium is reached.

\begin{figure}[htb]
\begin{center}
  \includegraphics[trim = 0 0 0 0, clip, width=\linewidth]{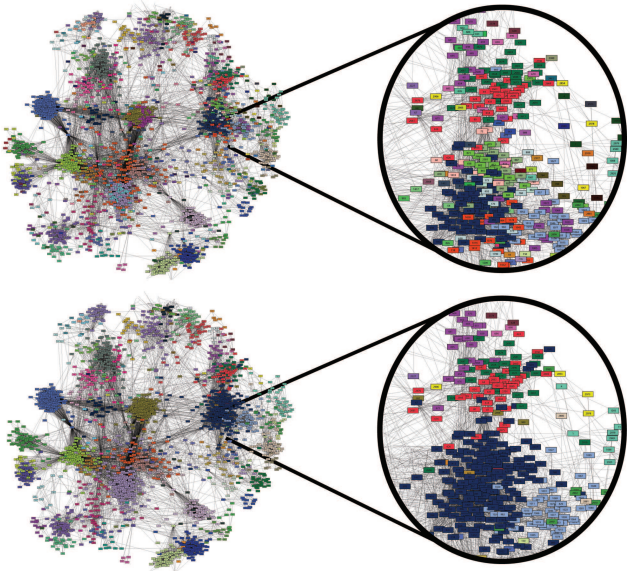}
  \caption{(Color online) Community detection in the OSN (n=3000) by gradually decreasing hop attenuation ($\delta = 0.5$ at the \emph{top} with $Q=0.64$, $\delta = 0$ at the \emph{bottom} with $Q=0.78$). Nodes with 3 or less neighbours are filtered to ease the visualisation.}\label{hierar}
\end{center} \end{figure}

Fig. \ref{hierar} gives an illustration of the first modification applied on a subgraph on the OSN. Note that this modification depends very much on the initial labelling of nodes because it determines the initial centers of these small communities.

Another important question which was also briefly addressed in \cite{linear} is the problem of overlapping communities \cite{k-cliqueCommunity}, i.e., nodes can often be considered a member of different communities. From previous sections, we understood that different asynchronous version of the algorithm is capable of generating very different results in different runs. This is exactly how \cite{linear} suggested as a potential solution - to re-run the algorithm several times. In a parallel environment, however, the results tend to be much less fluctuating. An initial attempt was to increase the number of labels passed each time between nodes to achieve a similar effect. Preliminary experiments indicate limited success since this setting hampers the convergence process, possibly due to the potential of latent labels switching back and fro in the system. Another possibility is the exploit the fact that nodes on the border of its community have different proportions (purity) of neighbours from other communities. We can potentially use that as a measure of membership but this indeed may only be applicable to such boundary nodes.

\subsection{Optimization}
The individual inspection of every node, particularly those with many neighbours, is a crucial factor in determining the speed of the algorithm. Putting aside efficient data structures and prudent programming, an obvious optimization we can do without much compromise on the performance is to selectively update high degree nodes. The reader may have realised that, after certain iterations, it would be pointless to update certain nodes that are well inside a cluster. These nodes are surrounded by nodes with the same label, which are unlikely to change for the same reason. We employ a simple purity measure of neighbours to selectively update nodes that are on the borders of their communities. In other words, we only update nodes whose number of neighbours sharing the maximal label is less than a certain percentage. Indeed, small degree nodes are likely to be avoided in early iterations in this setting but their contributions to the overall community structure and performance are almost insignificant. We carry out the modified algorithm with thresholds set at 100\% (equivalent to the unmodified algorithm), 80\%, 60\% and 40\% to examine the trade off between accuracy and speed.

\begin{figure}
\begin{center}
  \includegraphics[width=\linewidth]{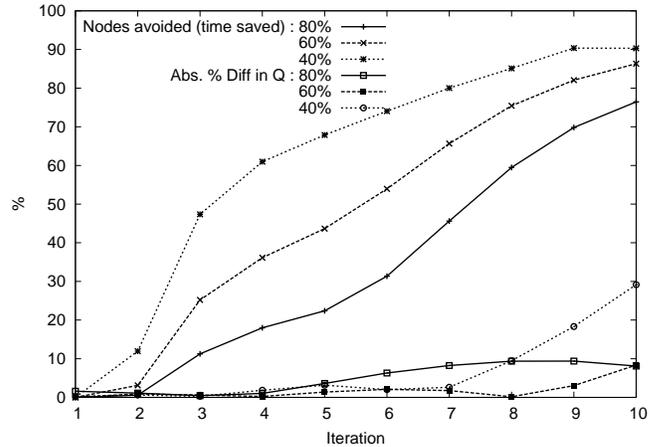}
  \caption{\label{ratio}The difference in \% modularity and speed of the optimized modifications with the original.}
\end{center} \end{figure}

Figure \ref{ratio} reveals that after the 1$^{st}$ iteration, the extra constraint will increasingly avoid updating nodes. As more nodes settle in a more stable cluster, increasingly less amount of time will be required in an iteration. Interestingly, even with a threshold as low as 40\%, the absolute difference in modularity compared to the original setting is reasonably small; and we can see the overall running time can be significantly reduced.

\subsection{Parallel \& Online Analysis}
Clear advantages of label propagation include its ease to be parallelized and its potential online implementation in real time networks. Since each node is required only to know information about its neighbours and updates itself according to the common rules, parallelism can be easily achieved. This brings us to another technical point that when the algorithm is completely parallelized, even without explicit synchronization, it would tend to behave like the synchronous version of the algorithm. And this is the key reason why we have stressed in the literature that improving both synchronous and asynchronous versions of the algorithm are equally important.

The running time in a parallel environment effectively reduces to $k$ if there are $\Theta(n)$ machines. This can be achieved in real world ubiquitous system such as a mobile ad-hoc network (MANET) or potentially on OSNs themselves (if members are willing to contribute their computational power) in real time. For instance, social information such as the community structure is known to benefit routing in MANET \cite{bubble}. Moreover, in such scenarios the space requirement for storing link information would become decentralised and thus insignificant.

On the same note, we see great potential in adapting the algorithm for online community detection in real-time dynamic networks where the presence of nodes and edges are constantly evolving. The microscopic movements and intermittent presence of nodes contribute to changes in terms of weights of the edges. These in turn result in five distinct macroscopic behaviours of communities, namely: growth, shrinkage, union, division and death of communities. The challenge indeed is to detect local changes without the need for a global update given limited computational resource or time constraint. We believe label propagation is particularly suited in this paradigm and thus propose this as future work.

\section{Comparisons}\label{comp}
We first look at two relatively large and previously studied networks for comparisons. These networks are respectively the Amazon Purchasing Network analysed in \cite{clauset-2004-70} and the actor collaboration network \cite{bbs}. As done in \cite{clauset-2004-70}, we assume all edges to be undirected to ease the analysis. With the added heuristics, the algorithm is able to perform within 5\% of CNM and 10\% of the adaptation by Danon, D\'{i}az-Guilera and Arenas (CNM-DDA) \cite{dda} in terms of modularity (c.f. Table \ref{tb}). LPA, however, achieves the result in a matter of minutes which is unparalleled by the above.

\begin{table*}[htb]
 \begin{center}
 \begin{tabular}
 {c|c|c|c|c|c}
 \hline \hline

 Network & Size & Directed Links & $Q$(Claimed) & Peak $Q$(Sync.) & Peak $Q$(Async.) \\ [0.5ex] \hline
 Amazon Purchase(Mar'03) & 409,687 & 4,929,260 & 0.745 \cite{clauset-2004-70} & 0.724 & 0.727 \\
 Actor Collaboration  & 374,511 & 30,052,912 & 0.528 \cite{linear}, 0.719 \cite{dda} & 0.642 & 0.660\\ \hline
 \end{tabular}
 \caption{The results correspond to the peak modularity achieved in 10 iterations or less, with  $f = Deg$ and $m = 0.1$ and a gradually decreasing $\delta$ as discussed in Section \ref{hie}.\label{tb}}
 \end{center}
\end{table*}

For a more standardized comparison, we turn to the recently proposed benchmark graphs by Lancichinetti et. al. \cite{bm}, an extension to the well known GN benchmark \cite{newman04findCommunity} which incorporates more realistic scale-free degree and cluster-size distributions. We follow closely the implementation of the benchmark graphs as described in \cite{bm} and compare the original LPA with the improved version on the graphs of size 1000 and 5000. To contrast label propagation with general fast modularity maximisation algorithms, we also run the benchmarks on the CNM algorithm.

\begin{figure*}[htb]
\centering
\includegraphics[trim = 90 230 70 0, clip, width=0.3\linewidth]{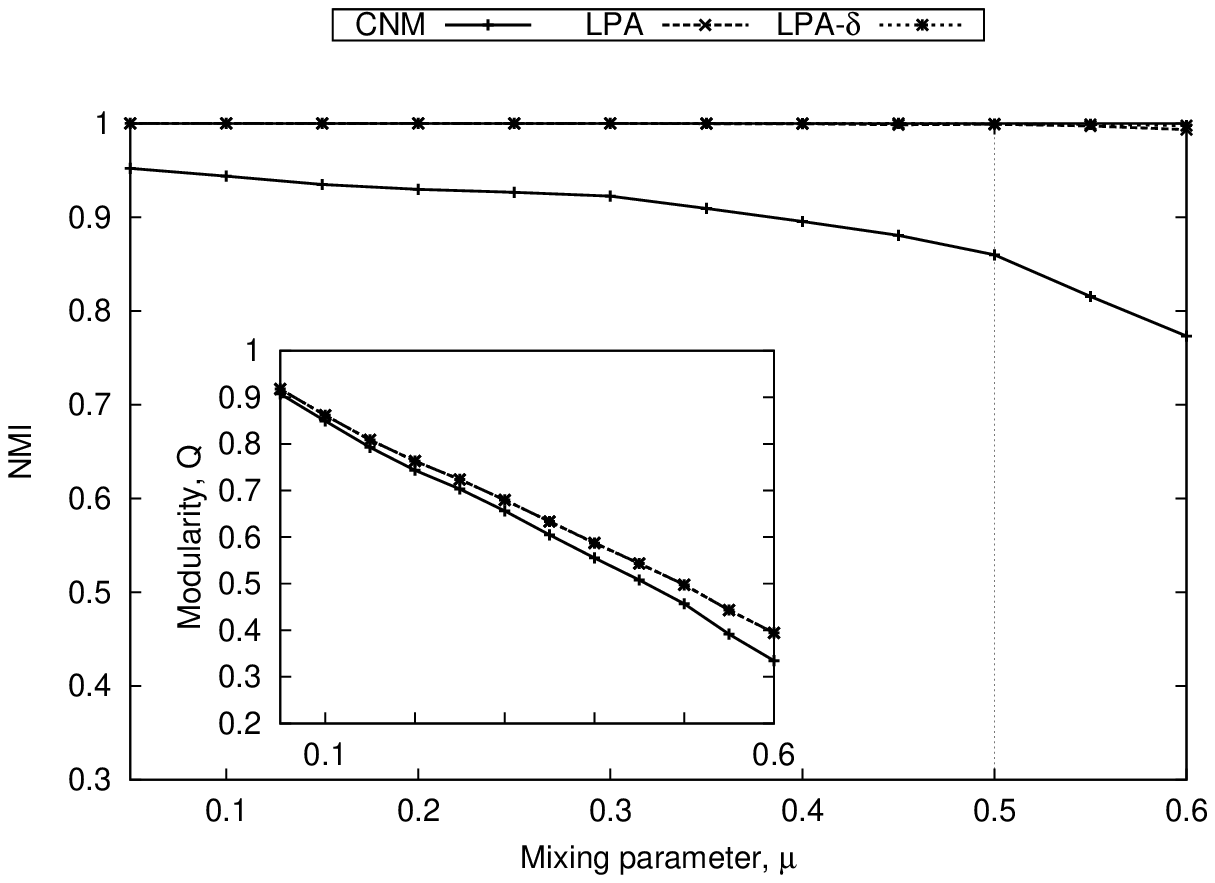}\\
    \subfigure[$N = 1000, \bar{d} = 15$]{\label{1k-15}\includegraphics[width=0.49\linewidth]{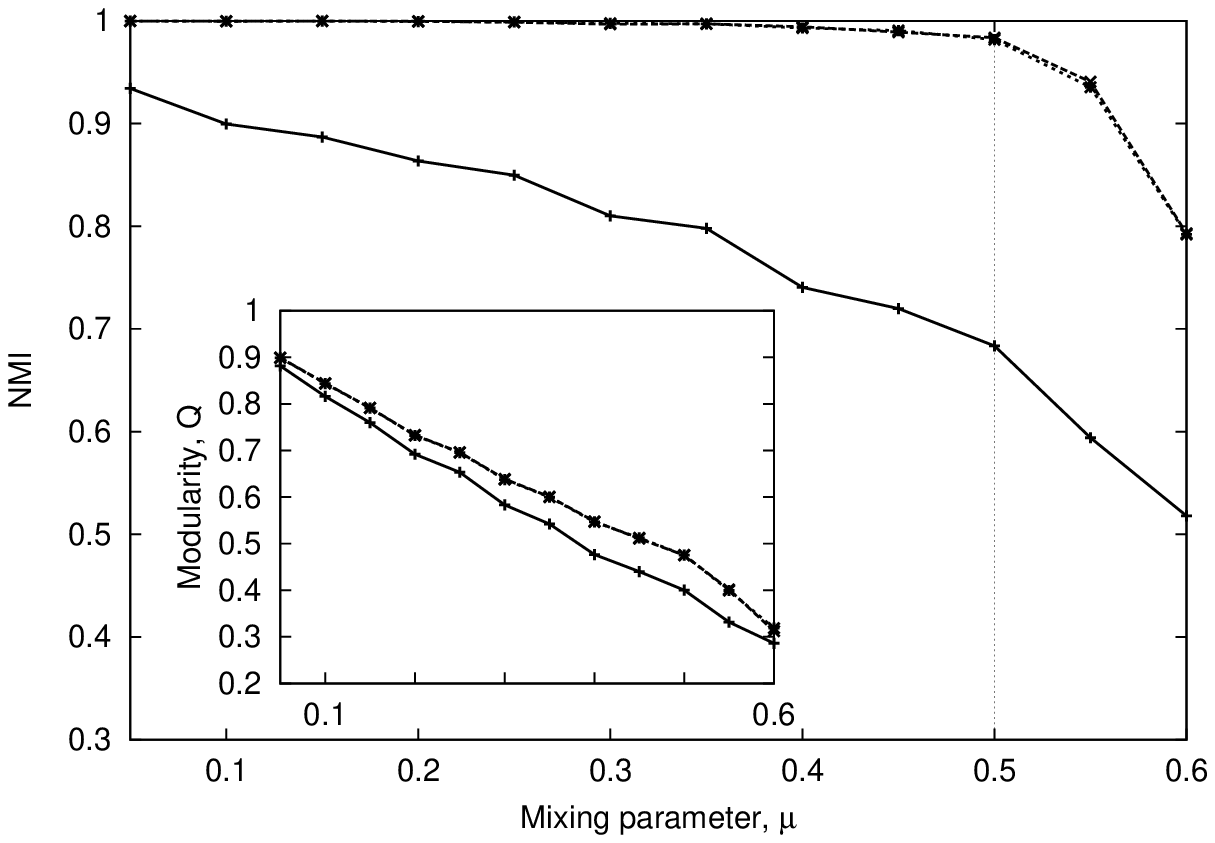}}
    \subfigure[$N = 1000, \bar{d} = 50$]{\label{1k-50}\includegraphics[width=0.49\linewidth]{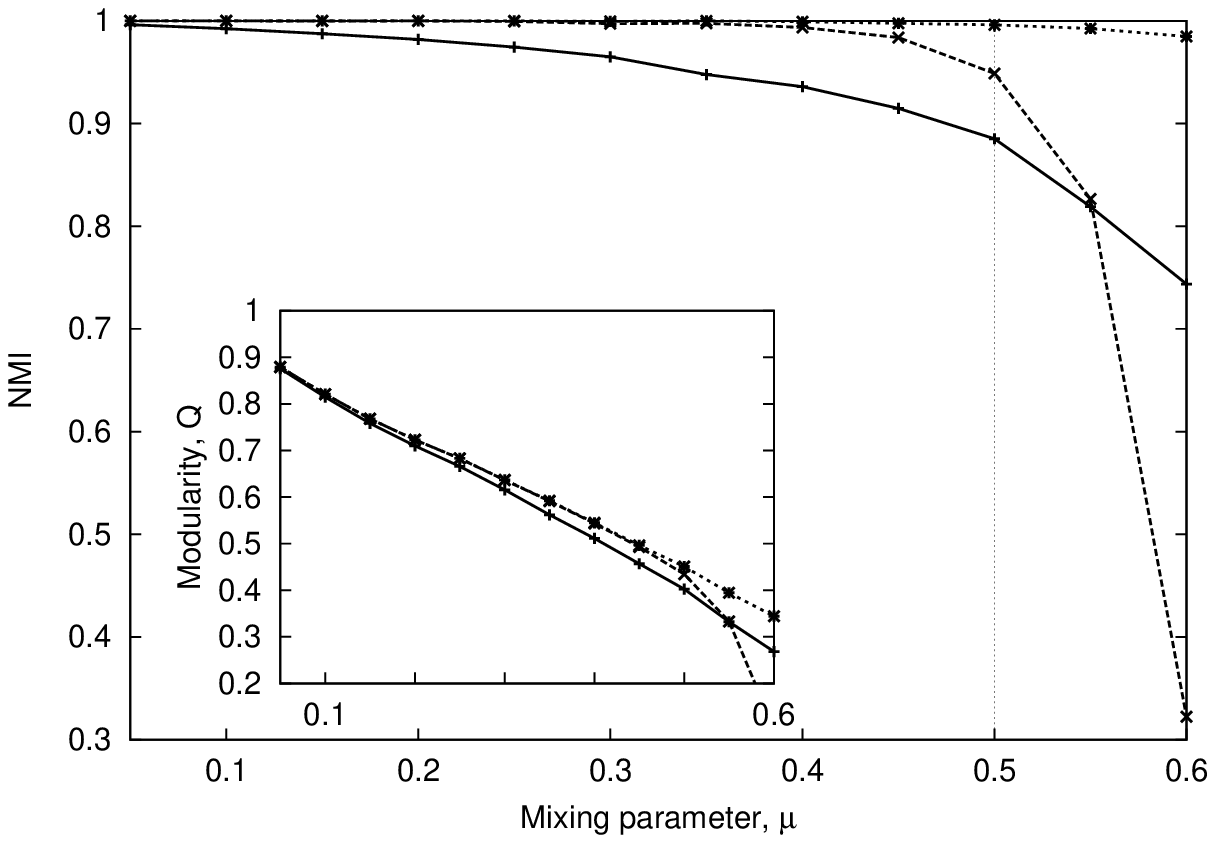}}\\
    \subfigure[$N = 5000, \bar{d} = 15$]{\label{5k-15}\includegraphics[width=0.49\linewidth]{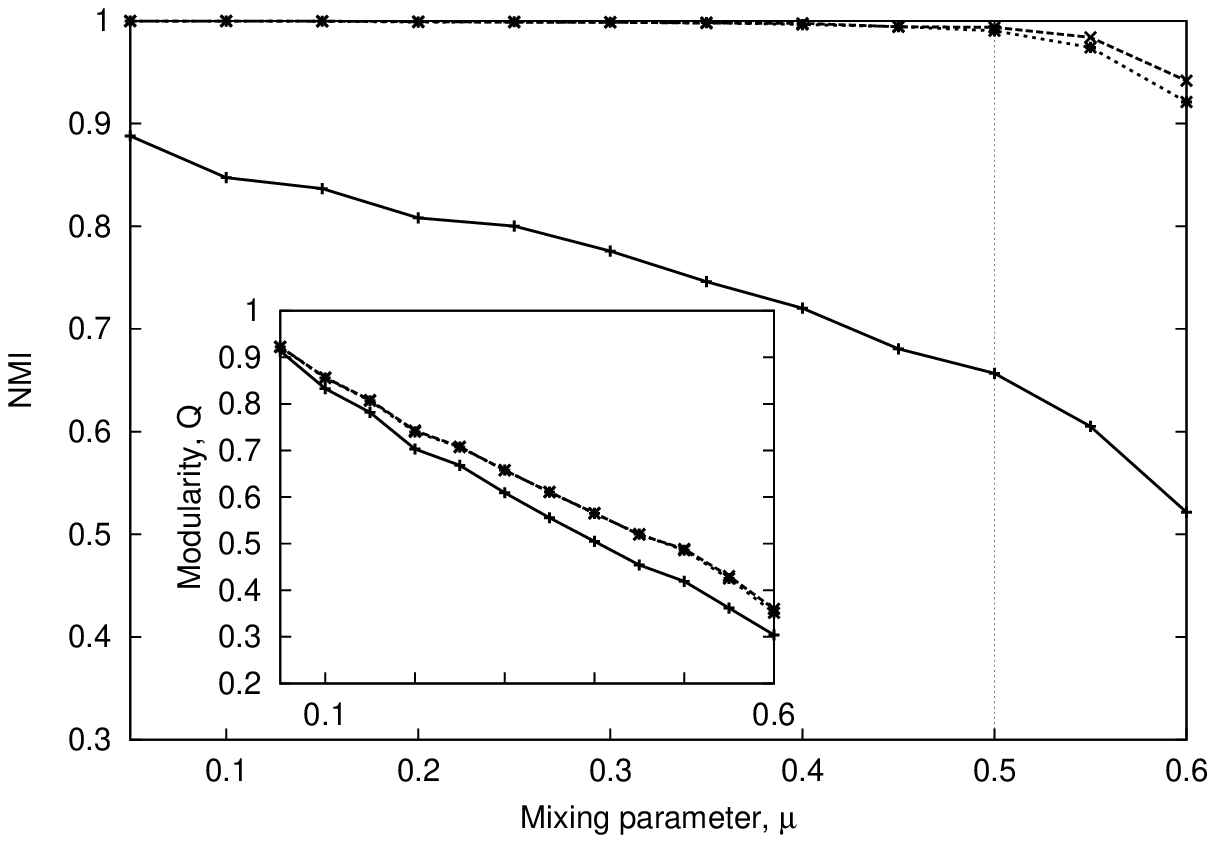}}
    \subfigure[$N = 5000, \bar{d} = 50$]{\label{5k-50}\includegraphics[width=0.49\linewidth]{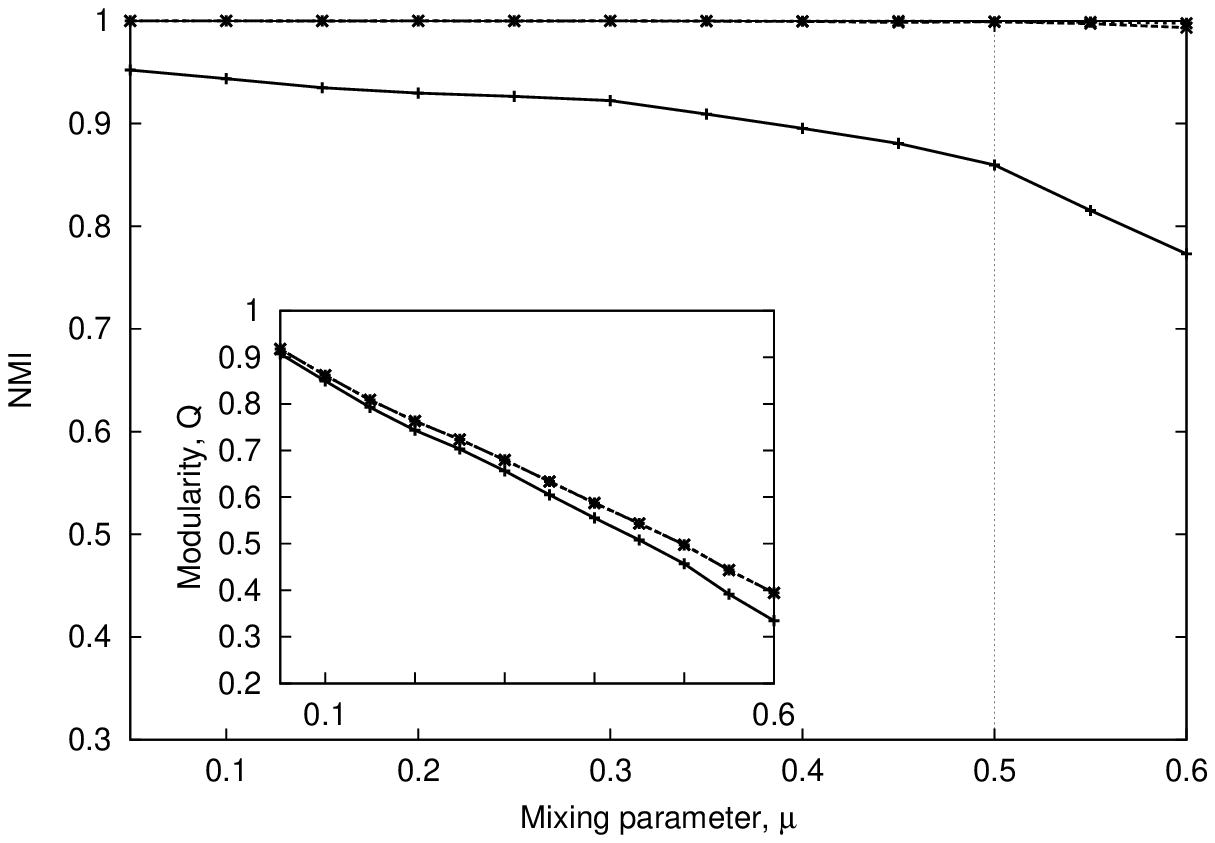}}
    \caption{Average performance comparisons between the three algorithms in the benchmark graphs with size $N$ and average degree $\bar{d}$. Both versions of LPA here are asynchronous; LPA-$\delta$ implements a gradually decreasing $\delta$ as discussed in Section \ref{hie}. All benchmark graphs have power-law degree and cluster-size distributions with exponent 3 and 2. For $N=1000$, the results are the average over 100 realisations; for $N=5000$, over 10 realisations.}
    \label{bmcomp}
\end{figure*}

As shown in Fig. \ref{bmcomp}, both implementations achieve superior accuracy over CNM in terms of normalised mutual information (NMI) even up to a mixing parameter of 0.6. Interestingly, the original method shows signs of failure at $\mu = 0.5$ in the $N = 1000, \bar{d} = 50$ benchmark graphs (c.f. Fig. \ref{1k-50}). We believe this corresponds to the formation of monster communities discussed in Section \ref{hop}. The number of nodes and the average degree of the benchmark graphs in effect dictate the number and sizes of the original communities generated. The results hence point out that denser and less modularized graphs are relatively prone to the formation of monster communities. However, the application of hop attenuation as exemplified in Fig. \ref{1k-50} greatly improves the overall performance of LPA in such scenarios.

Importantly, as opposed to label propagation, we can see that CNM algorithm's performance does not merely depend on the mixing parameter but also the average degree of the network. Resolution limit of modularity maximization is reflected by CNM's worse performance in graphs having a smaller average degree. Although in most configurations all algorithms expectedly manage to uncover a modularity value of a similar magnitude, the real accuracy in terms of NMI does not follow. This finding corresponds to the notion in \cite{reslimit} that modularity maximisation does not simply translate to actual communities.

\section{Conclusions}\label{con}
In this literature, we have empirically analysed a scalable, efficient and accurate community detection algorithm. We discussed the behaviours and emphasized the importance of both the synchronous and asynchronous implementations of the algorithm. We suggested potential heuristics that can be applied to improve its average detection performance and adaptability. Most importantly, we contrasted the algorithm with modularity-gain based methods in terms of community detection accuracy and observed how it can be potentially applied online and concurrently in large-scale and real-time dynamic networks.

Understanding the dynamics of this algorithm would be the major future work of this discipline before one devises further heuristics to improve the algorithm. We believe that each notion discussed in Section \ref{imp} is worthy of further inspection. An equally important point is to analyse mathematically or empirically on how to best adapt the algorithm to different types of networks by the added heuristics. How do different network topologies and models affect the algorithm's convergent behaviour? These are all valuable questions to be investigated in future work.

In summary, we show that label propagation with the appropriate modifications is a more reliable and efficient method in detecting communities in large-scale networks than popular existing methods. We trust that with further understanding and analysis epidemic-based community detection would be of substantial value to the field.

\begin{acknowledgments}
We thank Franco Bagnoli and Vito Latora for helpful comments. We are grateful to Eric Promislow for providing us with the Amazon network data. Network visualisations are carried out on Cytoscape \footnote{http://www.cytoscape.org/}. This project is supported by EC IST SOCIALNETS - Grant agreement number 217141.
\end{acknowledgments}

\bibliography{main}

\end{document}